\documentclass[twocolumn,prd,aps,nofootinbib,showpacs]{revtex4}
\usepackage{multirow}
\usepackage{graphicx}
\usepackage[dvips]{color}
\usepackage{amssymb,amsmath}

\begin{document}

\title{Lattice QCD calculation of $\pi\pi$ scattering length}
\author{Ziwen Fu}
\affiliation{
Key Laboratory of Radiation Physics and Technology {\rm (Sichuan University)},
Ministry of Education; \\
Institute of Nuclear Science and Technology, Sichuan University,
Chengdu 610064, P. R. China.
}

\begin{abstract}
We study $s$-wave pion-pion ($\pi\pi$) scattering length in lattice QCD
for pion masses ranging from $330$ MeV to $466$ MeV.
In the ``Asqtad'' improved staggered fermion formulation,
we calculate the $\pi\pi$ four-point functions for isospin $I=0$ and $2$ channels,
and use chiral perturbation theory at next-to-leading order
to extrapolate our simulation results.
Extrapolating to the physical pion mass gives the scattering lengths as
$m_\pi a_0^{I=2} = -0.0416(2)$ and  $m_\pi  a_0^{I=0} = 0.186(2)$
for isospin $I=2$ and $0$ channels, respectively.
Our lattice simulation for $\pi\pi$ scattering length
in the $I=0$ channel is an exploratory study,
where we include the disconnected contribution,
and our preliminary result is near to its experimental value.
These simulations are performed with MILC $2+1$ flavor gauge configurations
at lattice spacing $a \approx 0.15$~fm.
\end{abstract}


\maketitle

\section{Introduction}
Pion-pion scattering at low energies is elemental and important
hadron-hadron scattering process.
The $s$-wave $\pi\pi$ scattering lengths are predicted
at leading order (LO) in chiral perturbation theory ($\chi$PT)
by Weinberg~\cite{Weinberg:1966kf} in terms of pion mass,
$m_\pi$, and pion decay constant, $f_\pi$, as
$$
m_\pi a_{\pi\pi}^{I=0} \approx  \frac{7m_\pi^2}{16\pi f_\pi^2} = 0.160 ;
m_\pi a_{\pi\pi}^{I=2} \approx -\frac{m_\pi^2 }{ 8\pi f_\pi^2} = -0.0456 .
$$
The next-to-leading order (NLO)
corrections depend on unknown low energy constants, which can be
obtained from experimental measurements or lattice QCD.

The recent measurements of the $K_{e4}$ decays~\cite{Batley:2007zz} and
$K^{\pm}\rightarrow\pi^{\pm}\pi^0\pi^0$ decays~\cite{Batley:2000zz} by NA48/2 at
CERN~\cite{NA48} give, $m_\pi a_{\pi\pi}^{I=0}=0.221(5)$ and
$m_\pi a_{\pi\pi}^{I=2}=-0.0429(47)$.
Including $\chi$PT constraints in their analysis,
the determination of $s$-wave $\pi\pi$ scattering lengths
reaches~\cite{Colangelo:2001df,Leutwyler:2006qq}:
\begin{equation}
m_\pi a_{\pi\pi}^{I=0} =  0.220(5) ; \quad
m_\pi a_{\pi\pi}^{I=2} = -0.0444(10) .
\label{eq:roy}
\end{equation}

Lattice calculations of $\pi\pi$ scattering have been studied
in quenched QCD by various groups~\cite{Sharpe:1992pp,Kuramashi:1993ka,
Fukugita:1994na,Fukugita:1994ve,Liu:2001ss,Li:2007ey},
and first full QCD calculation of $\pi\pi$ scattering length
was done to study isospin $I=2$ $s$-wave scattering~\cite{Yamazaki:2004qb}.
First fully-dynamical calculation of $I=2$ $\pi\pi$ scattering length
was performed by NPLQCD~\cite{Beane:2005rj,Beane:2007xs}.
Mixed-action $\chi$PT at NLO was used to
perform the chiral and continuum extrapolations, and obtain
$$
m_\pi a_{\pi\pi}^{I=2} = -0.04330(42) \quad \textmd{and} \quad
{l_{\pi\pi}^{I=2}}(\mu) = 6.2 \pm 1.2 ,
$$
where ${l_{\pi\pi}^{I=2}}(\mu)$ is a low energy constant (LEC), which is
evaluated at physical pion decay constant $f_{\pi,\mathrm{phy}}$.
Using the $N_f=2$ maximally twisted mass fermion ensembles,
Xu Feng et al~\cite{Feng:2009ij} adopt the lightest pion mass, perform an explicit
check for large lattice artifacts, and find
$$
m_\pi {l_{\pi\pi}^{I=2}} = -0.04385(28), \quad
{l_{\pi\pi}^{I=2}}(\mu=f_{\pi,\mathrm{phy}})=4.65(85),
$$
which is in agreement with the above experimental measurements
and phenomenological analysis.

So far, only few efforts have been taken  for $I=0$ case.
Using the quenched approximation Y. Kuramashi et~al. explored the $I=0$ channel,
but the disconnect diagram was neglected for some reasons~\cite{Kuramashi:1993ka}.
Qi Liu performed  full QCD calculation
for the $I=0$  channel with the consideration of disconnected graph,
however the scattering length has a large error, and serves only
as a bound on the magnitude~\cite{Liu:2009uw}.

It is well-known that $I=0$ channel is a great challenging
phenomenologically because of $\sigma$ resonance.
Encouraged by our trustful measurement of
$\pi K$ scattering~\cite{Fu:2011wc},
here we use MILC gauge configurations generated
with $2+1$ flavors of Asqtad improved staggered
dynamical sea quarks~\cite{Bernard:2010fr} to investigate
$s$-wave $\pi\pi$  scattering for $I=0$ and $2$ channels.
We measure all the diagrams with extra efforts
to the disconnect diagram.
We note an attractive signal for the $I=0$ channel
and repulsive one in the $I=2$ case.
As presented later, after chiral extrapolation,
we find at the physical pion mass
$$
m_\pi a_{\pi\pi}^{I=0} =  0.186(2) ; \quad
m_\pi a_{\pi\pi}^{I=2} =  -0.0416(2) ,
$$
which is in reasonable agreement with above experimental measurements and
phenomenological analysis as well as previous lattice calculations.
Moreover, we perform an exploratory work for calculating ${l_{\pi\pi}^{I=0}}(\mu)$,
which is a LEC appearing in $\chi$PT description of the quark mass dependence of
the scattering length for the $I=0$ channel.

\section{Method}
\label{sec_method}
Let us study the scattering of two Nambu-Goldstone
pions  in the Asqtad-improved staggered dynamical fermion formalism
at zero momentum.
Here we follow the original conventions and notations
in Refs.~\cite{Sharpe:1992pp,Kuramashi:1993ka,Fukugita:1994ve}.
Using operators $O_\pi(x_1)$, $O_\pi(x_2)$ for pions at points $x_1$ and $x_2$, respectively,
we then express the $\pi\pi$ four-point functions as
$$
C_{\pi\pi}(x_4,x_3,x_2,x_1) =
\bigl< O_\pi(x_4) O_{\pi}(x_3) O_\pi^{\dag}(x_2) O_{\pi}^{\dag}(x_1)\bigr> ,
$$
where $\langle \cdots\rangle$ stands for the expectation value of path integral.
After summing over the spatial coordinates,
we achieve $\pi\pi$ four-point function in zero-momentum state,
\begin{equation}
\label{EQ:4point_pK}
C_{\pi\pi}(t_4,t_3,t_2,t_1) =
\sum_{{\bf{x}}_1} \sum_{{\bf{x}}_2} \sum_{{\bf{x}}_3} \sum_{{\bf{x}}_4}
C_{\pi\pi}(x_4,x_3,x_2,x_1) , \nonumber
\end{equation}
where $x_1 \equiv ({\textbf{x}}_1, t_1)$,
      $x_2 \equiv ({\textbf{x}}_2, t_2)$,
      $x_3 \equiv ({\textbf{x}}_3, t_3)$,
      $x_4 \equiv ({\textbf{x}}_4, t_4)$, and
$t$ represents time difference, i.e., $t\equiv t_3 - t_1$.
We choose $t_1 =0, t_2=1, t_3=t$, and $t_4 = t+1$
to avert the Fierz rearrangement of the quark lines,
and build  $\pi\pi$ operators
for isospin $I=0$ and $2$ eigenstates as
\begin{eqnarray}
\label{EQ:op_pipi}
 {\cal O}_{\pi\pi}^{I=0} (t) &=&  \frac{1}{\sqrt{3}}
     \Bigl\{  \pi^{-}(t) \pi^{+}(t+1) + \pi^{+}(t) \pi^{-}(t+1) - \cr
              &&\pi^{0}(t) \pi^{0}(t+1) \Bigl\} , \cr
 {\cal O}_{\pi\pi}^{I=2} (t) &=& \pi^{+}(t) \pi^{+}(t+1)  ,
\end{eqnarray}
with the pion interpolating field operators
\begin{eqnarray}
 \pi^+(t) &=&
 -\sum_{{\bf{x}}}\bar{d}({\bf{ x}}, t)\gamma_5 u({\bf{ x}},t) , \cr
\pi^-(t) &=& \sum_{{\bf{x}}} \bar{u}({\bf{x}},t)\gamma_5 d({\bf{x}},t) , \cr
\pi^0(t) &=&
\frac{1}{\sqrt{2}}\sum_{\bf{x}} [
 \bar{u}({\bf{x}}, t) \gamma_5 u({\bf{x}}, t) -
 \bar{u}({\bf{x}}, t) \gamma_5 d({\bf{x}}, t) ] . \nonumber
\end{eqnarray}

\begin{figure}[thb]
\includegraphics[width=8.0cm]{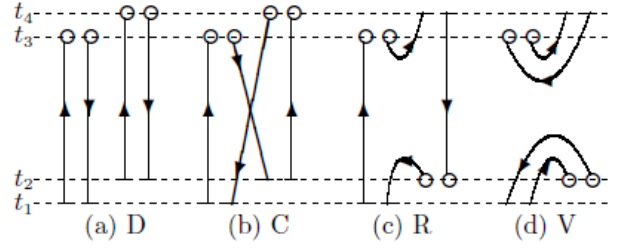}
\caption{ \label{fig:diagram}
Diagrams contributing to $\pi\pi$ four-point functions.
Short bars stand for wall sources.
Open circles are sinks for local pion operator.
}
\end{figure}

In the isospin limit, four diagrams contribute to $\pi\pi$ scattering amplitudes~\cite{Sharpe:1992pp,Kuramashi:1993ka,Fukugita:1994ve}.
The quark line diagrams contributing to $\pi\pi$ four-point function
are displayed in Figure~\ref{fig:diagram},
and labeled as direct($D$), crossed ($C$), rectangular ($R$),
and vacuum ($V$) diagrams, respectively.

It is very difficult to reliably measure the rectangular ($R$) and vacuum diagrams ($V$)~\cite{Kuramashi:1993ka,Fukugita:1994ve}.
In this work we use moving wall source method to solve
it~\cite{Fu:2011wc,Kuramashi:1993ka,Fukugita:1994ve}, namely,
each propagator, which corresponds
to a wall source at time slice
$ t = 0, \cdots, T-1$, are defined by
\begin{equation}
\sum_{n''}D_{n',n''}G_t(n'') =
\sum_{\textit{\textbf{ x}}}
\delta_{n',({\textit{\hspace{-0.10cm}\textbf{ x}}},t)}, \quad 0 \leq t \leq T-1 ,
\end{equation}
where $D$ is the quark matrix.
The combination of $G_t(n)$ that we apply for $\pi\pi$ four-point
functions is schematically shown in Figure~\ref{fig:diagram}.
and we can also express them in terms of the quark propagators $G$, namely,
\begin{widetext}
\begin{eqnarray}
\label{eq:dcr}
C_{\pi\pi}^D(t_4,t_3,t_2,t_1) &=&
\sum_{ \textit{\textbf{x}}_3} \sum_{ \textit{\textbf{x}}_4 }
\langle \mbox{Re} \, \mbox{Tr}
[G_{t_1}^{\dag}({\textit{\textbf{x}}_3}, t_3) G_{t_1}({\textit{\textbf{x}}_3}, t_3)
 G_{t_2}^{\dag}({\textit{\textbf{x}}_4}, t_4) G_{t_2}({\textit{\textbf{x}}_4}, t_4) ] \rangle,\cr
C_{\pi\pi}^C(t_4,t_3,t_2,t_1) &=&
\sum_{ \textit{\textbf{x}}_3} \sum_{ \textit{\textbf{x}}_4 }
\langle \mbox{Re} \, \mbox{Tr}
[G_{t_1}^{\dag}({\textit{\textbf{x}}_3}, t_3) G_{t_2}({\textit{\textbf{x}}_3}, t_3)
 G_{t_2}^{\dag}({\textit{\textbf{x}}_4}, t_4) G_{t_1}({\textit{\textbf{x}}_4}, t_4) ] \rangle ,\cr
C_{\pi\pi}^R(t_4,t_3,t_2,t_1) &=&
\sum_{ \textit{\textbf{x}}_2} \sum_{ \textit{\textbf{x}}_3 }
\langle \mbox{Re} \, \mbox{Tr}
[G_{t_1}^{\dag}({\textit{\textbf{x}}_2}, t_2)
 G_{t_4}({\textit{\textbf{x}}_2}, t_2)
 G_{t_4}^{\dag}({\textit{\textbf{x}}_3}, t_3)
 G_{t_1}({\textit{\textbf{x}}_3}, t_3) ] \rangle , \cr
 C_{\pi\pi}^V(t_4,t_3,t_2,t_1) &=&
\sum_{\textit{\textbf{x}}_2} \sum_{ \textit{\textbf{x}}_3}
\Bigl\{ \mbox{Re} \langle \mbox{Tr}
[G_{t_1}^{\dag}({\textit{\textbf{x}}_2}, t_2)
 G_{t_1}({\textit{\textbf{x}}_2}, t_2) \rangle
 \langle  \mbox{Tr} G_{t_4}^{\dag}({\textit{\textbf{x}}_3}, t_3)
 G_{t_4}({\textit{\textbf{x}}_3}, t_3) ]
\rangle  -\nonumber \\
&&\hspace{1.3cm}
 \mbox{Re} \langle \mbox{Tr}
[G_{t_1}^{\dag}({\textit{\textbf{x}}_2}, t_2)
 G_{t_1}({\textit{\textbf{x}}_2}, t_2)] \rangle
\langle  \mbox{Tr}
[G_{t_4}^{\dag} ({\textit{\textbf{x}}_3}, t_3)
 G_{t_4}({\textit{\textbf{x}}_3}, t_3) ]
\rangle \Bigl\} ,
\end{eqnarray}
\end{widetext}
where daggers is the conjugation by the even-odd parity $(-1)^n$
for the Kogut-Susskind quark action.
We utilize the hermiticity properties of the quark propagator $G$
to remove the factors of  $\gamma^5$.
The vacuum diagram here is accompanied by a vacuum subtraction~\cite{Blum:2011pu}.
As it is discussed in Refs.~\cite{Kuramashi:1993ka,Fukugita:1994ve},
the rectangular and vacuum diagrams create gauge-variant noise,
which are suppressed by performing the gauge field average
without gauge fixing~\cite{Kuramashi:1993ka,Fukugita:1994ve}
in the current study.

All the four diagrams in Figure~\ref{fig:diagram} can be combined to
build physical correlation functions
for $\pi\pi$ states with definite isospin.
If we consider that $u$ and $d$ quarks have the same mass,
the $\pi\pi$ correlation function for $I=0$ and $I=2$
can be written in terms of these diagrams, namely,
\begin{eqnarray}
\label{EQ:phy_I0I2}
C_{\pi\pi}^{I=0}(t) &\equiv&
\left\langle {\cal O}_{\pi\pi}^{I=0} (t) | {\cal O}_{\pi\pi}^{I=0} (0)
\right\rangle   \cr
&=& D + \frac{N_f}{2}C - 3N_f R + \frac{3}{2}V  , \cr
C_{\pi\pi}^{I=2}(t) &\equiv&
\left\langle {\cal O}_{\pi\pi}^{I=2} (t) | {\cal O}_{\pi\pi}^{I=2} (0) \right\rangle
= D - N_f C ,
\end{eqnarray}
where $N_f$ is inserted to address the flavor degrees of freedom of
the Kogut-Susskind staggered fermion~\cite{Sharpe:1992pp}.

The $s$-wave $\pi\pi$ scattering length in the continuum is defined by
\begin{equation}
\label{eq:exact}
a_0 = \lim_{k\to 0} \frac{\tan\delta_0(k)}{k} .
\end{equation}
$k$ is the magnitude of the center-of-mass scattering momentum
related to the total energy by
$E_{\pi\pi}^I = 2\sqrt{m_\pi^2 + k^2}$ of the $\pi\pi$ system
in a cubic box of size $L$ with isospin $I$.
$\delta_0(k)$ is $s$-wave scattering phase shift,
which can be calculated by the L\"uscher's formula~\cite{Luscher:1991p2480,Lellouch:2001p4241},
\begin{equation}
\left( \frac{\tan\delta_0(k)}{k} \right)^{-1} =
\frac{\sqrt{4\pi}}{\pi L} \cdot {\mathcal Z}_{00}\left(1,\frac{k^2}{(2\pi/L)^2}\right) ,
\label{eq:luscher}
\end{equation}
where the zeta function $\mathcal{Z}_{00}(1;q^2)$ is defined by
\begin{equation}
\label{eq:Z00d}
\mathcal{Z}_{00}(1;q^2) = \frac{1}{\sqrt{4\pi}}\sum_{{\mathbf n}\in\mathbb{Z}^3} \frac{1}{n^2-q^2}\;,
\end{equation}
here $q = kL/(2\pi)$, and $\mathcal{Z}_{00}(1;q^2)$
can be efficiently computed the method discussed in Refs.~\cite{Yamazaki:2004qb,Fu:2011wc}.

The energy $E_{\pi\pi}$ can be obtained from
the $\pi\pi$ four-point function
denoted in Eq.~(\ref{EQ:phy_I0I2}) with the large $t$.
At large  $t$ this correlator will fall as
\begin{equation}
\label{eq:E_pionpion}
C_{\pi\pi}^I(t) \propto  e^{-E_{\pi\pi} t} + \cdots ,
\end{equation}
where $E_{\pi\pi}$ is the energy of the lightest two pion state.
In the usual manner, pion mass $m_\pi$ can be evaluated through
\begin{equation}
\label{eq:E_pion}
C_\pi(t)  \propto   e^{-m_\pi t} + \cdots .
\end{equation}

In our concrete calculation we evaluate the energy shift
$\delta E = E -  2m_\pi$ from the ratio
\begin{equation}
\label{EQ:ratio}
R^X(t) \hspace{-0.02cm}=\hspace{-0.02cm} \frac{ C_{\pi\pi}^X(0,1,t,t+1) }
       { C_\pi (0,t) C_\pi(1,t+1) }, \quad  X\hspace{-0.05cm}=D, C, R, \ {\rm and} \ V , \nonumber
\end{equation}
where $C_\pi (0,t)$ and $C_\pi (1,t+1)$ are the pion two-point functions.
The amplitudes which project out the $I=0$ and $2$
isospin eigenstates  can be written as
\begin{eqnarray}
\label{EQ:proj_I0I2}
R_{I=0}(t) \hspace{-0.2cm}&=& \hspace{-0.2cm}
R^D(t) + \frac{N_f}{2}R^C(t) - 3N_f R^R(t) + \frac{3}{2}R^V(t), \cr
R_{I=2}(t) \hspace{-0.2cm}&=& \hspace{-0.2cm}
R^D(t) - N_f R^C(t) .
\end{eqnarray}
We extract the energy shift $\delta E$ from the ratio~\cite{Sharpe:1992pp}
\begin{equation}
\label{eq:extraction_dE}
R_I(t) \hspace{-0.05cm}=\hspace{-0.05cm} Z_I e^{-\delta E t} + \cdots ,
\end{equation}
where $Z_I$ stands for wave function factor~\cite{Kuramashi:1993ka,Fukugita:1994ve}.

\section{Simulation results}
We use the MILC lattices in the presence of the $2+1$ dynamical flavors
of the Asqtad-improved staggered dynamical fermions,
the description of its simulation parameters
are given in Refs.~\cite{Bernard:2010fr,Bazavov:2009bb}.
We analyzed the $\pi\pi$ four-point functions on the $0.15$ fm MILC ensemble
of $360$ $16^3 \times 48$ gauge configurations
with bare quark masses $am_{ud}/am_s = 0.0097/0.0484$
and bare gauge coupling $10/g^2 = 6.572$,
which has an inverse lattice spacing $a^{-1}=1.358^{+35}_{-12}$ GeV~\cite{Bernard:2010fr,Bazavov:2009bb}.
The masses of the $u$ and $d$ quarks are degenerate.

We use the standard conjugate gradient method to obtain
the required matrix element of the inverse fermion matrix.
We compute the propagators on
all the time slices $ t = 0, \cdots, T-1$ of both source and sink.
After averaging the correlators over all $T=48$ possible values,
the statistics are significantly improved
because we can place the pion source at all time slices.

With same configurations we compute the $\pi\pi$ four point correlation
functions using six $u$ valence quarks, namely,
$am_x = 0.0097$, $0.01067$, $0.01261$, $0.01358$, $0.01455$, and $0.0194$,
where $m_x$ is the light valence $u$ quark mass.

\begin{figure}[tbh]
\includegraphics[width=8cm,clip]{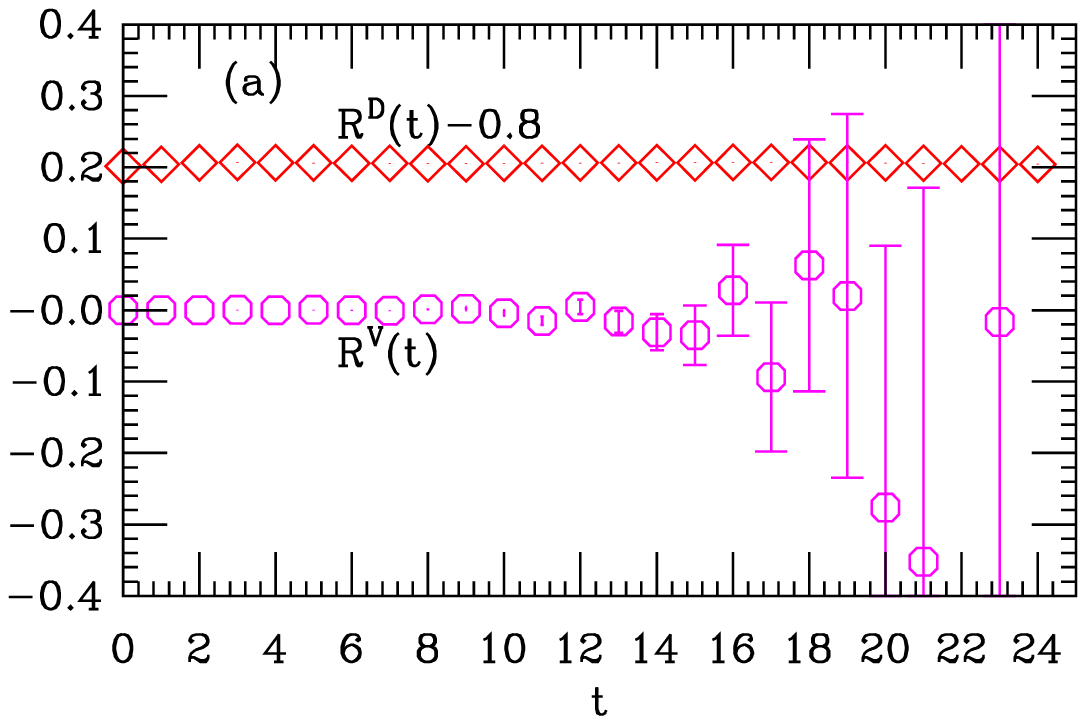}
\includegraphics[width=8cm,clip]{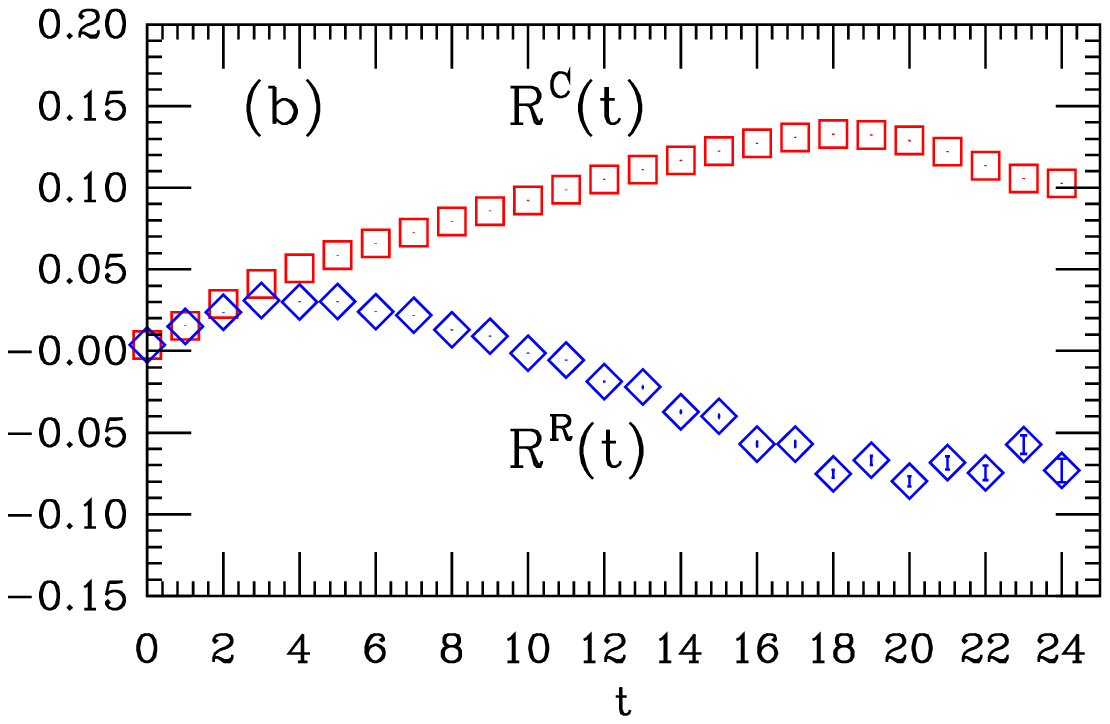}
\caption{\label{fig:ratio}
Individual amplitude ratios $R^X(t)$ as functions of $t$.
(a) Direct diagram shifted by $0.8$ (diamonds) and vacuum diagram (octagons);
(b) crossed (squares) and rectangular (diamonds) diagrams.
}
\end{figure}

In Figure~\ref{fig:ratio} the individual ratios,
which are defined in Eq.~(\ref{EQ:ratio}) corresponding to the diagrams
in Figure~\ref{fig:diagram}, $R^X$ ($X=D, C, R$ and $V$)
are displayed as functions of $t$ for $am_x=0.0097$.
It is extremely noisy for the disconnected diagram($V$),
but still we can get a good signal up to time separation $t=14$.
Clear signals observed up to $t = 19$ for the rectangular
amplitude and up until $t = 14$ for the vacuum amplitude
demonstrate that the method of wall source without gauge fixing
used here is practically applicable.

The values of the direct amplitude $R^D$ is quite close to unity,
indicating a weak interaction in this channel.
The crossed amplitude, on the other hand,
increases linearly up to $t \sim 17$,
implying a repulsion in the $I=2$ channel.
After a beginning increase up to $t \sim 4$,
the rectangular amplitude
shows a linear decrease up until $t \sim 20$,
suggesting an attractive force between two pions.
Moreover, the magnitude of this slope is analogous  to
that of the cross amplitude but with different sign.
These characteristics are
what we want~\cite{Weinberg:1966kf,Sharpe:1992pp}.

The vacuum amplitude is negligibly small up to $t \sim 10-14$,
and loss of signals after that.
This characteristic is in well accordance with the Okubo-Zweig-Iizuka (OZI) rule
and $\chi$PT in leading order, which expects the disappearing
of the vacuum amplitude~\cite{Kuramashi:1993ka}.
Moreover, its the errors should be approximately
independent of time separation $t$,
and increases exponentially like $\displaystyle e^{2m_\pi t}$~\cite{Kuramashi:1993ka}.
Therefore, it is very difficult to obtain its proper information
from large time separation.

\begin{figure}[tbh]
\includegraphics[width=8cm,clip]{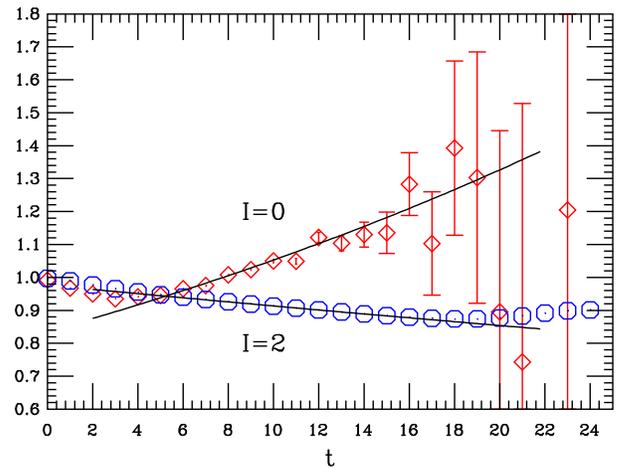}
\caption{\label{fig:I0I2}
$R_I(t)$ for $\pi\pi$ four-point function
calculated without gauge fixing for $am_x=0.0097$.
Solid line in $I=2$ is exponential fits for $10 \le t \le 15$ and
Solid line in $I=0$ is exponential fits for $7 \le t \le 14$.
}
\end{figure}

In Figure~\ref{fig:I0I2} we plot the ratio $R_I(r)$ projected
onto the isospin $I = 0$ and $2$ channels for $am_x=0.0097$,
which are denoted in Eq.~(\ref{EQ:proj_I0I2}).
A decrease of the ratio of $R_{I=2}(t)$ indicates  a positive energy shift
and hence a repulsive interaction for the $I = 2$ channel,
and an increase of $R_{I=0}(t)$ suggests an attraction for the $I = 0$ case.
A dip at $t=3$ for the $I = 0$ channel can be clearly noted~\cite{Fukugita:1994ve}.

In this work, we use Eq.~(\ref{eq:extraction_dE})
to extract the energy shift $\delta E_I$,
and then insert them into
the Eqs.~(\ref{eq:exact}) and (\ref{eq:luscher}) to
obtain the scattering lengths.
In this work, the energy shifts $a\delta E$
are picked up from the effective energy shift plots,
and they were selected by searching for
a ``plateau'' in the energy shift as a function of the minimum distance $\rm D_{min}$ as well as a good fit quality (namely, $\chi^2$)~\cite{Fu:2011wc}.

We utilize the exponential physical fitting model in Eq.~(\ref{eq:extraction_dE})
to extract the desired energy shifts for both $I=2$ and $0$ channels.
In Figure~\ref{fig:I0I2} we display the ratio $R_I(t)$ projected
onto both channels for $am_x=0.0097$.
The fitted values of the energy shifts, $\delta E_I$ in lattice units,
fitting range,
and wave function factor $Z_I$ 
are summarized in Table~\ref{tab:energy_shifts}.
The third block shows energy shifts in lattice units,
Column four shows wave function factor $Z_I$,
Column five shows time range for the chosen fit,
and Column six shows degrees of freedom (dof) for the fit.
The wave function $Z_I$ factors are pretty close to unity
and the $\chi^2/{\rm dof}$ is pretty small for the $I=2$ channel,
indicating the values of the extracted scattering lengths are substantially reliable,
and the $Z$ factors are also near to unity,
and the $\chi^2/{\rm dof}$ is reasonable for the $I=0$ channel,
suggesting the value of the extracted scattering lengths are enough safe.

\begin{table}[h!]
\caption{\label{tab:energy_shifts}
Summary of simulation results for energy shifts.
}
\begin{ruledtabular}
\begin{tabular*}{80mm}{cllllc}
I & $m_x$   & $\delta E$ & $Z_I$ & Range  & $\chi^2/{\rm dof}$ \\
\hline
\multirow{6}*{$0$}
&$0.00970$  & $-0.0241(19)$   & $0.825(7) $ & $7-14$ & $12.8/6$  \\
&$0.01067$  & $-0.0235(32)$   & $0.835(23)$ & $7-14$ & $9.46/6$ \\
&$0.01261$  & $-0.0222(29)$   & $0.854(19)$ & $6-12$ & $6.94/5$ \\
&$0.01358$  & $-0.0223(30)$   & $0.856(20)$ & $6-12$ & $5.33/5$ \\
&$0.01455$  & $-0.0217(32)$   & $0.863(20)$ & $6-10$ & $4.66/3$ \\
&$0.01940$  & $-0.0199(30)$   & $0.884(19)$ & $6-12$ & $16.5/5$ \\
\hline
\multirow{6}*{$2$}
&$0.00970$  & $0.00670(15)$   & $0.977(2)$  & $10-15$ & $1.07/4$ \\
&$0.01067$  & $0.00662(15)$   & $0.980(2)$  & $10-15$ & $0.17/4$ \\
&$0.01261$  & $0.00648(14)$   & $0.984(2)$  & $10-15$ & $0.05/4$ \\
&$0.01358$  & $0.00640(14)$   & $0.986(2)$  & $10-15$ & $0.03/4$ \\
&$0.01455$  & $0.00625(13)$   & $0.986(2)$  & $10-15$ & $0.02/4$ \\
&$0.01940$  & $0.00594(12)$   & $0.993(2)$  & $10-15$ & $0.08/4$ \\
\end{tabular*}
\end{ruledtabular}
\end{table}

In our previous work~\cite{fzw:2011cpl,Fu:2011wc},
we have measured the pion masses ($m_\pi$) and the pion decay constants $f_\pi$~\cite{Aubin:2004fs},
which are summarized in Table~\ref{tab:f_pi}.
The second and third blocks show pion masses in lattice unit and in GeV,
respectively, and Column four shows the pion decay constants in lattice units.
\begin{table}[h!]
\caption{\label{tab:f_pi}
Summary of the pion mass and the pion decay constants.
}
\begin{ruledtabular}
\begin{tabular*}{80mm}{c@{\extracolsep{\fill}}cccc}
$m_x$     & $a m_\pi$   & $m_\pi({\rm GeV})$ & $a f_{\pi}$  \\
\hline
$0.00970$ & $0.2458(2)$ & $0.334(6)$  & $0.12136(29)$ \\
$0.01067$ & $0.2575(2)$ & $0.350(6)$  & $0.12264(34)$ \\
$0.01261$ & $0.2787(2)$ & $0.379(7)$  & $0.12425(27)$ \\
$0.01358$ & $0.2890(2)$ & $0.392(7)$  & $0.12482(32)$ \\
$0.01455$ & $0.2987(2)$ & $0.406(7)$  & $0.12600(26)$ \\
$0.01940$ & $0.3430(2)$ & $0.466(8)$  & $0.12979(27)$ \\
\end{tabular*}
\end{ruledtabular}
\end{table}
Now we can substitute  these energy shifts in Table~\ref{tab:energy_shifts}
into the Eq.~(\ref{eq:exact})
to achieve the scattering lengths.
The center-of-mass scattering momentum $k^2$ in GeV calculated by
$E_{\pi K}^I = \delta E_I + 2m_\pi = 2\sqrt{m_\pi^2 + k^2}$
and then its $s$-wave scattering lengths $a_0$ obtained through Eq.~(\ref{eq:exact})
for both $I=0$, and $2$ channels are summarized in Table~\ref{tab:I2I0}.
The errors come from the statistic errors of
the fitted values of the energy shifts.
The third block shows center-of-mass scattering momentum $k^2$ in GeV,
Column four shows $s$-wave scattering lengths in lattice units,
and Column five shows pion mass times $s$-wave scattering lengths.

\begin{table}[t]	
\caption{\label{tab:I2I0}
Summary of lattice simulation scattering lengths.
}
\begin{ruledtabular}
\begin{tabular*}{80mm}{cllll}
Isospin &$m_x$   &  $k^2$[GeV]  & $a_0$  & $m_\pi a_0$\\
\hline
\multirow{6}*{$0$}
&$0.00970$  & $-0.0107(4)$   & $2.95(19)$   & $0.724(48) $\\
&$0.01067$  & $-0.0109(7)$   & $3.03(32)$   & $0.781(85) $\\
&$0.01261$  & $-0.0112(7)$   & $3.16(33)$   & $0.882(92) $\\
&$0.01358$  & $-0.0117(8)$   & $3.40(38)$   & $0.983(107) $\\
&$0.01455$  & $-0.0118(8)$   & $3.67(38)$   & $1.095(112) $\\
&$0.01940$  & $-0.0124(9)$   & $3.78(50)$   & $1.297(173) $\\
\hline
\multirow{6}*{$2$}
&$0.00970$  & $0.00306(7)$  & $-0.495(5)$  &  $-0.121(2) $\\
&$0.01067$  & $0.00316(7)$  & $-0.509(5)$  &  $-0.131(1)$\\
&$0.01261$  & $0.00335(7)$  & $-0.537(5)$  &  $-0.150(1)$\\
&$0.01358$  & $0.00343(7)$  & $-0.548(5)$  &  $-0.159(2)$\\
&$0.01455$  & $0.00345(7)$  & $-0.552(5)$  &  $-0.165(2)$\\
&$0.01940$  & $0.00378(8)$  & $-0.598(6)$  &  $-0.205(2)$\\
\end{tabular*}
\end{ruledtabular}
\end{table}

We adopt the formula predicted by $\chi$PT at NLO
to extrapolate $\pi\pi$ scattering lengths to the physical point.
As suggested in Refs.~\cite{Beane:2005rj,Beane:2007xs,Feng:2009ij},
we carry out the chiral extrapolation of $m_\pi a^{I=2}_{\pi\pi}$ and
$m_\pi a^{I=0}_{\pi\pi}$  in terms of $m_\pi/f_\pi$ instead of $m_\pi$.
Thus we use the continuum $\chi$PT forms
of $a_0^{I=2}$ and $a_0^{I=0}$, which are directly constructed
from Appendix C in Ref.~\cite{Bijnens:1997vq}, as
\begin{widetext}
\begin{eqnarray}
\label{eq:fit}
\label{eq:m_pipi_I0}
m_\pi a^{I=0}_{\pi\pi}
 &=& \;\, \frac{7m_\pi^2}{16\pi f_\pi^2}
 \left\{1-\frac{m_\pi^2}{16\pi^2 f^2_\pi}
\left[9\ln \frac{m_\pi^2}{f_\pi^2}-
5 - l_{\pi\pi}^{I=0}(\mu=f_{\pi,\mathrm{phy}})
\right]\right\}, \\
\label{eq:m_pipi_I2}
m_\pi a^{I=2}_{\pi\pi}
 &=& -\frac{m_\pi^2}{8\pi f_\pi^2}\left\{1+\frac{m_\pi^2}{16\pi^2 f^2_\pi}
\left[3\ln \frac{m_\pi^2}{f_\pi^2}-1-
l_{\pi\pi}^{I=2}(\mu=f_{\pi,\mathrm{phy}})\right]\right\},
\end{eqnarray}
\end{widetext}
\noindent
where we plugged in the values of the pion mass $m_\pi$,
and the pion decay constants $f_\pi$,
which are summarized Table~\ref{tab:f_pi},
and $l_{\pi\pi}^{I=0}(\mu)$ and $l_{\pi\pi}^{I=2}(\mu)$
are related to the Gasser-Leutwyler
coefficients $\bar{l}_i$ as~\cite{Bijnens:1997vq}
\begin{eqnarray}
\label{eq:l_pipi_I0}
l_{\pi\pi}^{I=0}(\mu) \hspace{-0.1cm}&=&\hspace{-0.1cm}
\frac{40}{21}\bar{l}_1+\frac{80}{21}\bar{l}_2-\frac{5}{7}\bar{l}_3+4\bar{l}_4
+9\ln\frac{m_{\pi,\mathrm{phy}}^2}{{\mu^2}}\,, \\ 
\label{eq:l_pipi_I2}
l_{\pi\pi}^{I=2}(\mu) \hspace{-0.1cm}&=&\hspace{-0.1cm}
\frac{8}{3}\bar{l}_1+\frac{16}{3}\bar{l}_2-\bar{l}_3-4\bar{l}_4+
3\ln\frac{m_{\pi,\mathrm{phy}}^2}{{\mu^2}}\,,
\end{eqnarray}
here Eq.~(\ref{eq:l_pipi_I2}) were extensively used
in Refs.~\cite{Beane:2005rj,Beane:2007xs,Feng:2009ij}.

\begin{figure}[h]
\includegraphics[width=80mm]{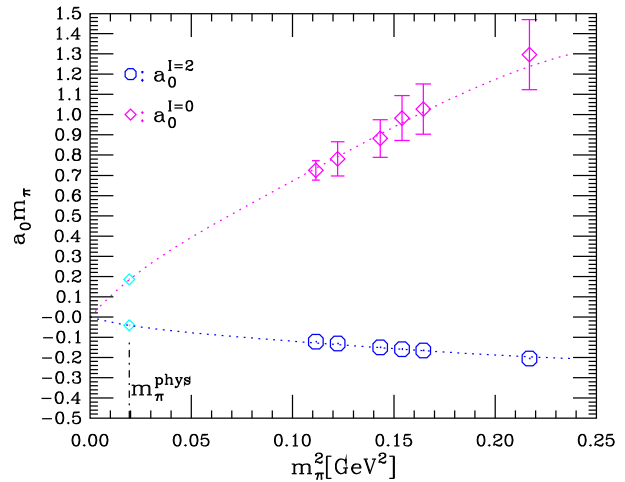}
\caption{\label{fig:ChPT-fit}
$m_{\pi}^2$-dependence of $\pi\pi$ scattering lengths $m_{\pi} a_0$
for $I=0$ and $2$ channels.
The dotted lines give $\chi$PT predictions at NLO.
The cyan diamond points indicate its physical values.
}
\end{figure}

The chiral extrapolation of $\pi\pi$ scattering lengths,
$m_\pi a_0^{I=2} $ and $m_\pi a_0^{I=0}$
are plotted by the dotted lines as a function of $m_\pi^2$ in Figure~\ref{fig:ChPT-fit},
The fit parameters $l_{\pi\pi}^{I=0}(\mu)$, $l_{\pi\pi}^{I=2}(\mu)$,
and the $s$-wave scattering lengths
$m_\pi a_0$ at the physical points,
where we adopt the latest PDG~\cite{Nakamura:2010zzi} values,
are also summarized in Table~\ref{tb:ChPT-fit}.
The chiral scale is taken as $\mu=f_{\pi,\mathrm{phy}}$.
From Figure~\ref{fig:ChPT-fit}, we can note that
our lattice simulation results for the $I=2$ scattering length
agrees well with the one-loop formula,
while scattering length for $I=0$ have a large error,
and is also in reasonable agreement with $\chi$PT at NLO.

\begin{table}[h!]
\caption{\label{tb:ChPT-fit}
The fitted $m_\pi a_0$ at the physical point.
}
\begin{ruledtabular}
\begin{tabular*}{80mm}{cccc}	
$\mathrm{Isospin}$ &$\chi^2/{\mathrm{dof}}$  &
$l_{\pi\pi}^I(\mu=f_{\pi,\mathrm{phy}})$     & $m_\pi a_0$   \\
\hline
$I=0$ &$ 0.268/5 $  & $ 18.674 \pm 1.213$  &   $0.186(2)$    \\
$I=2$ &$ 9.864/5 $  & $ 11.587 \pm 0.871$  &   $-0.0416(2)$  \\
\end{tabular*}
\end{ruledtabular}
\end{table}

Although the fitted value of $l_{\pi\pi}^{I=2}(\mu)$  is larger than
that of other lattice studies~\cite{Beane:2005rj,Beane:2007xs,Feng:2009ij},
our fitted value of $m_\pi a^{I=2}_{\pi\pi}$ is reasonable consistent with
other lattice studies~\cite{Beane:2005rj,Beane:2007xs,Feng:2009ij}.
Since it is an exploratory study for $I=0$ channel,
there are no lattice comparisons
with our fitted values of $l_{\pi\pi}^{I=0}(\mu)$ and $m_\pi a_0$.
Anyway, our fitted value of $s-$wave $\pi\pi$ scattering length for $I=0$ channel is
in reasonable agreement with the experimental measurement in Eq.~(\ref{eq:roy}).

\section{Conclusions and outlooks}
In this work, we performed a lattice study of
$s$-wave $\pi\pi$ scattering for isospin $I=0$ and $2$ channels.
We evaluated all of the four diagrams,
and observed an attractive signal for the $I=0$ channel and
an repulsive one for the $I=2$ channel, respectively.
Extrapolating toward the physical point
yields $m_\pi  a_{0}^{I=2} = -0.0416(2)$ and  $m_\pi a_{0}^{I=0} = 0.186(2)$
for $I=2$ and $0$ channels, respectively,
which are in good agreement with $\chi$PT at NLO,
and $m_\pi a_{0}^{I=2} = -0.0416(2)$ is reasonably
consistent with other lattice
studies~\cite{Beane:2005rj,Beane:2007xs,Feng:2009ij}.
Moreover, we give an exploratory fitted value of
$s-$wave $\pi\pi$ scattering length for $I=0$ channel,
which is in reasonable agreement with
the recent experimental measurement~\cite{NA48}.

It is quite stimulating that $\pi\pi$ scattering
for the $I=0$ channel can be trustworthily calculated
by wall sources without gauge fixing.
It gives us hope that we can use this technique
to tackle $\sigma$ resonance,
which is still poorly understood from lattice QCD.

In our previous works~\cite{Ph.D:2007fzw,fzw:2011cpl,fzw:2011cpc10},
we studied and evaluated $\sigma$ mass,
and found that the $\sigma$ meson is heavier than the $\pi\pi$
threshold for enough small $u$ quark mass.
These works as well as our lattice investigation
for the $\pi\pi$ scattering in the $I=0$ channel
will stimulate people to investigate the decay mode
$\sigma \to \pi\pi$.

Since our study is restricted only at zero momenta,
it can not provide adequate information on $\sigma$ meson.
To investigate the $\sigma$ resonance,
we should  study the $\pi\pi$ scattering length
at the $I=0$ channel with non-zero momentum,
which is an indicative of a $\sigma$ pole.
We are beginning this kind of work,
and the measurement of the $\pi\pi$ scattering
for the $I=0$ channel with the momentum $p=(1,0,0)$ is in progress.



\begin{thebibliography}{90}
\bibitem{Weinberg:1966kf} S.~Weinberg,
Phys. Rev. Lett. {\bf 17} (1966) 616.

\bibitem{Batley:2007zz} J.~R. Batley et~al,
Eur. Phys. J., C {\bf 54} (2008) 411.

\bibitem{Batley:2000zz} J.~R. Batley et~al,
Eur. Phys. J., C {\bf 64} (2009) 589.

\bibitem{NA48} Brigitte Bloch-Devaux,
PoS, KAON09 (2009) 033.

\bibitem{Leutwyler:2006qq} H.~Leutwyler,
arXiv:hep-ph/0612112.

\bibitem{Colangelo:2001df} G.~Colangelo, J.~Gasser, and H.~Leutwyler,
Nucl. Phys., B {\bf 603} (2001) 125.

\bibitem{Sharpe:1992pp} S.~R.~Sharpe, R.~Gupta and G.~W.~Kilcup,
Nucl. Phys. B {\bf 383} (1992) 309.

\bibitem{Kuramashi:1993ka} Y.~Kuramashi et al,
Phys. Rev. Lett.  {\bf 71} (1993) 2387.

\bibitem{Fukugita:1994na} M.~Fukugita et al,
Phys. Rev. Lett.  {\bf 73} (1994) 2176.

\bibitem{Fukugita:1994ve} M.~Fukugita et al,
Phys. Rev. D {\bf 52} (1995) 3003.

\bibitem{Liu:2001ss} C.~Liu, J.~h.~Zhang, Y.~Chen and J.~P.~Ma,
  Nucl. Phys. B {\bf 624} (2002) 360.

\bibitem{Li:2007ey} X.~Li {\it et al.},
  JHEP {\bf 0706} (2007) 053.

\bibitem{Yamazaki:2004qb} T.~Yamazaki et al,
Phys. Rev. D {\bf 70} (2004) 074513.

\bibitem{Beane:2005rj} S.~R.~Beane, P.~F.~Bedaque, K.~Orginos and M.~J.~Savage,
Phys. Rev. D {\bf 73} (2006) 054503.

\bibitem{Beane:2007xs} Silas~R. Beane et al,
Phys. Rev., D {\bf 77} (2008) 014505.

\bibitem{Feng:2009ij} X.~Feng, K.~Jansen, D.~B.~Renner,
Phys. Lett. B  {\bf 684} (2010) 268.

\bibitem{Liu:2009uw} Q.~Liu,
PoS LAT2009 (2009) 101.

\bibitem{Fu:2011wc} Z.~Fu,
Phys.\ Rev.\  D {\bf 85}, 074501 (2012). arXiv:1110.1422 [hep-lat]

\bibitem{Bernard:2010fr} C.~Bernard et al,
Phys. Rev. D {\bf 83} (2011) 034503.

\bibitem{Blum:2011pu} T.~Blum et al,
arXiv:1106.2714 [hep-lat].

\bibitem{Gupta:1990mr} R.~Gupta, G.~Guralnik, G.~W.~Kilcup, S.~R.~Sharpe,
Phys. Rev. D {\bf 43} (1991) 2003.

\bibitem{Luscher:1991p2480} M.~Luscher,
Nuclear Physics B {\bf 354} (1991) 531.

\bibitem{Lellouch:2001p4241} L.~Lellouch and M.~Luscher,
Commun. Math. Phys. {\bf 219} (2001) 31.

\bibitem{Bazavov:2009bb} A.~Bazavov et al,
Rev. Mod. Phys.  {\bf 82} (2010) 1349.

\bibitem{fzw:2011cpl} Z. Fu,
Chin. Phys. Lett. {\bf 28} (2011) 081202.

\bibitem{Aubin:2004fs} C.~Aubin et~al,
Phys. Rev. D {\bf 70} (2004) 114501.

\bibitem{Bijnens:1997vq} J.~Bijnens, G.~Colangelo, G.~Ecker, J.~Gasser, and M.~E. Sainio,
Nucl. Phys., B {\bf 508} (1997) 263.

\bibitem{Nakamura:2010zzi} Nakamura K et al,
J. Phys. G {\bf 37} (2010) 075021.

\bibitem{Ph.D:2007fzw} C. Bernard et al,
Phys. Rev. D, {\bf 76} (2007) 094504

\bibitem{fzw:2011cpc10} Z. Fu  and C. DeTar,
Chin. Phys. C {\bf 35}(10) (2011) 896.

\end{thebibliography}
\end{document}